\documentclass[pre, amsfonts, amssymb, amsmath, showkeys, nofootinbib, longbibliography, twocolumn]{revtex4-1}
\usepackage{graphicx}
\usepackage[dvipsnames]{xcolor}
\usepackage{lipsum}
\usepackage{csquotes}
\usepackage{siunitx}
\usepackage{ulem} 

\definecolor{darkblue}{rgb}{0,0,0.6}
\definecolor{darkred}{rgb}{0.6,0,0}
\usepackage[pdftex, pdftitle={Article}, pdfauthor={Author}, colorlinks=true,urlcolor=darkblue, citecolor=darkblue, linkcolor=darkred, hyperfootnotes=false]{hyperref}

\begin{document}
\newcommand{\ellbc}{\ell_\mathrm{BC}}
\newcommand{\ellbg}{\ell_\mathrm{BG}}
\newcommand{\lambdadef}{\lambda_\mathrm{def}}
\newcommand{\tW}{\tilde{W}}
\newcommand{\tWc}{\tilde{W}_c}
\newcommand{\srr}{\sigma_{rr}}
\newcommand{\sqq}{\sigma_{\theta\theta}}
\newcommand{\stt}{\sigma_{\theta\theta}}
\newcommand{\tDelta}{{\tilde{\Delta}}}
\newcommand{\tdelta}{{\tilde{\delta}}}
\newcommand{\tepsilon}{{\tilde{\epsilon}}}
\newcommand{\rmur}{{\rm u}_r}
\newcommand{\UFvK}{U_{\rm FvK}}
\newcommand{\Ustrain}{U_{\rm strain}}
\newcommand{\Ubend}{U_{\rm bend}}
\newcommand{\Uad}{U_{\rm ad}}
\newcommand{\bW}{\bar{W}}
\newcommand{\bL}{\bar{L}}
\newcommand{\bWc}{\bar{W}_c}
\newcommand{\varepsilontt}{\varepsilon_{\theta\theta}}
\newcommand{\varepsilonrr}{\varepsilon_{rr}}
\newcommand{\glv}{\gamma_\textrm{l-v}}

\title{Rucks and folds: delamination from a flat rigid substrate under uniaxial compression}

\author{Benny Davidovitch}
    \email{bdavidov@umass.edu}
    \affiliation{Department of Physics, University of Massachusetts Amherst, Amherst, MA 01003}
 
\author{Vincent D\'emery}
\email{vincent.demery@espci.psl.eu}
\affiliation{Gulliver, CNRS, ESPCI Paris PSL, 10 rue Vauquelin, 75005 Paris, France}
\affiliation{Univ Lyon, ENS de Lyon, Univ Claude Bernard Lyon 1, CNRS, Laboratoire de Physique, F-69342 Lyon, France}   


\begin{abstract}
We revisit the delamination of a solid adhesive sheet under uniaxial compression from a flat, rigid substrate.  
Using energetic considerations and scaling arguments we show that the phenomenology is governed by three dimensionless groups, which characterize the level of confinement imposed on the sheet, as well as its extensibility and bendability. Recognizing that delamination emerges through a subcritical bifurcation from a planar, uniformly compressed state, we predict that 
the dependence of the threshold confinement level on the extensibility and bendability of the sheet, as well as the delaminated shape at threshold, varies markedly between two asymptotic regimes of these parameters. For sheets whose bendability is sufficiently high the delaminated shape is a large-slope ``fold'', where the amplitude is proportional to the imposed confinement.
In contrast, for lower values of the bendability parameter the delaminated shape is a small-slope ``ruck'', whose amplitude increases more moderately upon increasing confinement. 
Realizing that the instability of the fully laminated state requires a finite extensibility of the sheet, we introduce a simple model that allows us to construct a bifurcation diagram that governs the delamination process.
\end{abstract}


\maketitle

\section{Introduction}

\begin{figure}
\begin{center}
\includegraphics[scale=1]{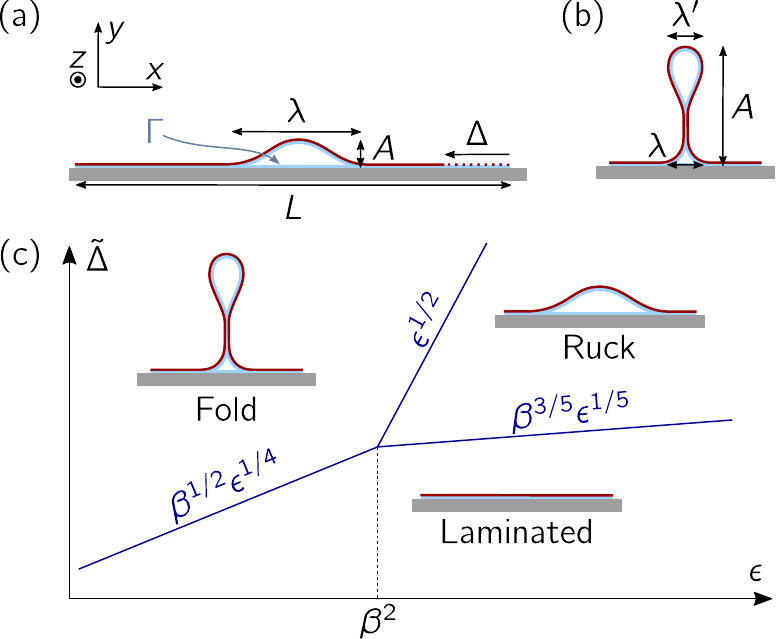}
\caption{(a) Schematic of our model system for delamination of a thin sheet under unaxial confinement ($\Delta$) from a flat rigid substrate.
(b) Fold ansatz.
(c) Schematic phase diagram, projected onto the plane of dimensionless parameters $(\epsilon, \tDelta)$ for a fixed value of $\beta$, showing the three possible morphologies: laminated, small-slope ruck and large-slope fold.
}
\label{fig:scheme_main}
\end{center}
\end{figure}

Beyond its broad importance for numerous branches of materials industry, such as coating and stretchable electronics, the delamination of thin sheets from an adhesive substrate provides an invaluable glance into the nontrivial mechanics of slender bodies \cite{Vella2019,Bico2018,Holmes2019,Paulsen2019}. 
This viewpoint motivated several groups to consider a thin sheet under uniaxial compression as an inextensible body, and employ Euler's {\emph{elastica}} to study its delamination from various types of adhesive substrates---a compliant solid \cite{Vella2009}, liquid bath~\cite{Wagner2011, Oshri2018}, or a flat rigid substrate \cite{Wagner2013}. 
The inherent simplicity of uniaxial compression, whereby the stress field in the sheet is a scalar function (often a constant), has also been exploited to study the gravity-limited deflection of a heavy sheet from a non-adhesive floor as a model for deformation patterns in rugs \cite{Vella2009,Kolinski2009}, and the periodic delamination pattern from an adhesive compliant substrate \cite{Vella2009}.
A related, yet far richer delamination morphology is observed when exerting a biaxial compression on an 
adhesive sheet that is supported on a rigid \cite{Gioia1997,Jagla2007} or compliant substrate \cite{Bedrossian2015,Hutchinson1992,Moon2004}, or upon attaching a sheet \cite{Majidi08} or a shell \cite{Bense2020} onto a substrate of spherical shape. The morphological complexity in such cases stems from the nontrivial relaxation of a non-uniform stress through delamination zones \cite{Hure11,Hure2013} and substrate deformation \cite{Hohlfeld15}.       
With an eye towards this rich behavior, we revisit in this article the problem of delamination of a nearly inextensible sheet from a rigid adhesive substrate induced by uniaxial compression (Fig.~\ref{fig:scheme_main}).

Wagner and Vella showed that Euler's elastica describes quantitatively the delaminated shape up to self-contact~\cite{Wagner2013}.
Notably they found that the energy $U$ of the delaminated state evolves with the length $\Delta$ absorbed in the blister as $U\sim\Delta^{1/3}$ for small $\Delta$.
However, their analysis does not predict the shape of the sheet after self-contact, which occurs at $\Delta\simeq 8\ellbc$, where $\ellbc$ is the bendo-capillary length~\cite{Paulsen2019}.
Moreover, under the inextensibility assumption, delamination occurs at $\Delta_\textrm{d}=0$ and requires an infinite compressive stress~\cite{Napoli2017}, similarly to the formation of a ruck in a rug~\cite{Lee2015Role}.

For a physical sheet, which can accomodate a finite, albeit minute level of compression, delamination is a discontinuous (subcritical) transition occurring at a finite compression length $\Delta_\textrm{d}>0$~\cite{Kolinski2009, Napoli2017}, where the amplitude ``jumps'' to a finite value.
Assuming that the delaminated shape is a small-slope ``ruck'', one can show that the sheet delaminates at $\Delta_\textrm{d}\ll\ellbc$, and find the dependence of $\Delta_\textrm{d}$ on the elastic moduli of the sheet and the adhesion energy~\cite{Kolinski2009}.
However, numerous swelling experiments of polymer films attached to rigid substrates \cite{Singamaneni2010,Ortiz2010,Velankar2012} indicate that patterns of large-slope, fold-like structures may emerge directly from a uniformly compressed state, rather than grow gradually from small-slope rucks (often described as ``buckles'').
While some works attributed these observations to a deformation localized at the film surface that involves a nontrivial stress profile across the film's thickness (similarly to creasing instabilities \cite{Tanaka1987}), it has been argued by Velankar {\emph{et al.}} that the emergence of folds in these experiments may reflect a delamination instability, describable by slender body mechanics and governed by balance of bending and adhesion energies~\cite{Velankar2012}.
However, to our knowledge, this proposal has not led to a quantitative theoretical analysis that addresses the validity of the assumption of small-slope at threshold, nor how $\Delta_\textrm{d}$ depends on the system parameters if the delaminated shape is instead a large-slope ``fold''.      

Here, we propose a large-slope ``fold'' ansatz for the shape of the sheet after self-contact.
By comparing this state to the laminated state and to a small-slope ruck, we determine the possible behaviors of a thin sheet under compression.
We find that moderately bendable sheets delaminate into a small-slope ruck, which then grows and finally turns into a large-slope fold---a scenario which has been noted previously \cite{Wagner2013}.
In contrast, highly bendable sheets delaminate directly into a large-slope fold (Fig.~\ref{fig:scheme_main}(b)).

This article is organized as follows.
In Sec.~\ref{sec:model_main} we define the model for wet adhesion on a rigid substrate and give our main results.
In Sec.~\ref{sec:Inextent} we perform a scaling analysis assuming inextensible delaminated shapes and derive our main result.
In Sec.~\ref{sec:Bifurcation} we relax the inextensibility constraint and study in detail the mechanics of a small-slope ruck and obtain its bifurcation diagram.
In Sec.~\ref{sec:defect} we study the defect mediated nucleation of delaminated states of small and large slope.
In Sec.~\ref{sec:beyond} we address the applicability of our results for delamination under uniaxial compression beyond the case of wet adhesion on a rigid substrate. Specifically, we consider dry adhesion and the effect of (conservative) tangential forces, gravity-limited delamination from a non-adhesive substrate, and delamination from a liquid bath.  
We discuss our results and conclude in Sec.~\ref{sec:discussion}.

\section{Model and main result}
\label{sec:model_main}

\subsection{Model}\label{}

We consider a sheet of length $L$, stretching modulus $Y=Et$, where $E$ is the Young's modulus and $t$ the thickness of the sheet, and bending modulus $B\sim Et^3$, in contact with a planar substrate at $y=0$ (Fig.~\ref{fig:scheme_main}(a)).
We assume that the system is invariant in the direction $z$, so that the shape of the sheet can be described by a curve in the $(x,y)$ plane.
Finally, the sheet is compressed along the $x$ axis by bringing its two ends together by a distance $\Delta$.

We assume that the sheet adheres to the substrate through a thin liquid film that wets the substrate and the sheet, such that both of them remain covered with a layer of the liquid film after any portion of the sheet detaches from the substrate.
The adhesion energy penalty is associated with the liquid-vapor interfaces, with surface energy $\glv$, which are created when the sheet is not in contact with the substrate or with itself.
The characteristic surface energy per area is thus $\Gamma=2\glv$.

The energy of our model thus consists of three terms: the adhesion, bending, and strain energies,
\begin{equation}
U =\Uad +  \Ubend + \Ustrain . \label{eq:Utot}
\end{equation}

From the parameters $Y$, $B$, $\Gamma$, $L$ and $\Delta$, we can construct three dimensionless groups:
\begin{align}
\beta & = \frac{\Gamma}{Y},\\
\epsilon & = \frac{B}{\Gamma L^2} = \left(\frac{\ellbc}{L} \right)^2, \label{eq:epsilon}\\
\tDelta & = \frac{\Delta}{L}.
\end{align}
Here, $\beta$ is the ``extensibility'' (\emph{i.e.} adhesion-induced strain), $\epsilon^{-1}$ is the ``bendability'', $\tDelta$ is the confinement, and $\ellbc$ is the bendo-capillary length:
\begin{equation}\label{eq:lbc}
\ellbc=\sqrt{\frac{B}{\Gamma}}.
\end{equation}
Here we focus on the limit of a nearly inextensible, highly bendable sheet, $\beta\ll 1$, $\epsilon\ll 1$.

\begin{figure}
\begin{center}
\includegraphics[scale=1]{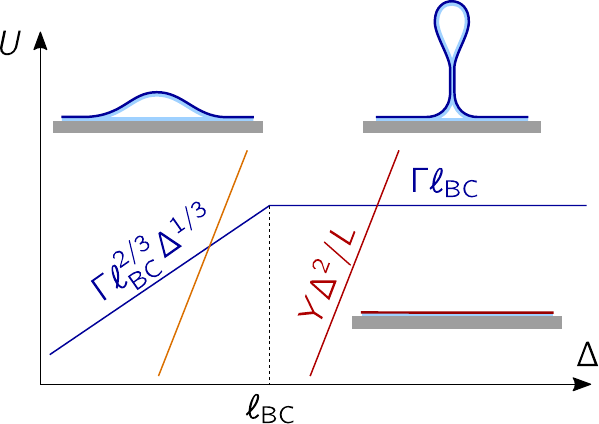}
\caption{Schematic log-log plot of the energy of the various states upon increasing $\Delta$. Blue: small-slope ruck followed by large-slope fold. Orange and red: laminated state, for two different values of $L$. The orange line corresponds to the moderate bendability regime, the red line corresponds to the high bendability regime.
}
\label{fig:Energy}
\end{center}
\end{figure}

The laminated state contains only strain energy: its energy per unit length in the transverse direction $z$ is (Fig.~\ref{fig:Energy})
\begin{equation}\label{eq:Ulam}
U^\mathrm{lam}=\frac{Y\Delta^2}{2L}=\frac{YL\tDelta^2}{2}.
\end{equation}

\subsection{Main result}\label{}

Our main prediction, shown in Fig.~\ref{fig:scheme_main}, is the splitting of the asymptotic regime into two domains for the critical confinement where delamination occur, $\tDelta_\textrm{d}$:
\begin{itemize}
\item[(a)] High bendability regime: 
\begin{equation}
\epsilon \ll \beta^2 : \quad\tDelta_\textrm{d}\sim \beta^{1/2}  \epsilon^{1/4}  
\label{eq:VHB}
\end{equation} 
in which the energetically-favorable state at $\tDelta \!> \!\tDelta(\epsilon,\beta)$ is a large-slope ``fold'', 
whose amplitude is proportional to the exerted confinement ($A\sim \tDelta$). 
\item[(b)] Moderate bendability regime: 
\begin{equation}
\beta^2 \ll \epsilon : \quad\tDelta_\textrm{d}\sim \beta^{3/5}  \epsilon^{1/5}  
\label{eq:MHB}
\end{equation} 
in which the energetically-favorable state at $\tDelta > \tDelta_\textrm{d}$ is a small-slope ``ruck'', 
whose amplitude increases more gradually with the exerted confinement ($A\sim \tDelta^{2/3}$), until the ruck becomes a self-touching, large-slope fold at $\tDelta^* \sim \epsilon^{1/2} \gg \tDelta_\textrm{d}$ (corresponding to $\Delta\sim\ellbc$).
\end{itemize}

In order to explore the relevance of our predictions for actual physical systems, it is useful to express the inequality~(\ref{eq:VHB}) that defines the high bendability regime in terms of the physical parameters: 
\begin{equation}
{\rm high\ bendability:} \ \ \frac{L}{t} \gg \left(\frac{Et}{\Gamma}\right)^{{3}/{2}}\gg 1 \ ,   
\label{eq:22}
\end{equation} 
(we ignored $O(1)$ numerical prefactors).
For an ultrathin spin-coated sheet of Polystyrene or PMMA ($E \sim 1~\si{\giga \pascal}$), with thickness $t \in (10-1000)~\si{\nano m}$, of the type used in many elasto-capillary experiments \cite{Paulsen2019} and a characteristic value of the adhesion energy, $\Gamma \approx 0.1~\si{N/m}$, we find that the high bendability regime may be realized by sheets whose length $L$ is a few micrometers or more.

\section{Scaling analysis of compressed lamination versus compression-free delamination}
\label{sec:Inextent}

In this section we further simplify the analysis by assuming that the strain in the laminated state is completely relieved at the delamination transition, such that the delaminated state satisfies $\Ustrain = 0$, and can thus be described by the Euler's {\emph{elastica}}.
Moreover, we restrict ourselves to a scaling analysis.
In Sec.~\ref{sec:Bifurcation} we will elaborate on the validity of this assumption.

\subsection{Small-slope ruck}
\label{sub:scaling_ruck}

Following Refs.~\cite{Kolinski2009, Napoli2017}, we start by assuming that the delaminated state is a small-slope ruck, whose amplitude $A$ is significantly smaller than its width $\lambda$ (Fig.~\ref{fig:scheme_main}(a)).
Since we assume that there is no strain in the delaminated state ($\Ustrain = 0$), the inward displacement $\Delta$ must be absorbed in the excess length of the ruck, leading to the ``slaving'' condition
\begin{equation} \frac{A^2}{\lambda} \sim \Delta.
\label{eq:slaving_ruck}
\end{equation}

The adhesion and bending energies of the ruck then result in
\begin{align}\label{eq:Uruck}
U^\mathrm{ruck}&\sim \Gamma\lambda+\frac{BA^2}{\lambda^3}\\
&\sim \Gamma\lambda+\frac{B\Delta}{\lambda^2},\label{eq:Uruck2}
\end{align}
where the curvature of the sheet is $\sim A/\lambda^2$, and we have used the slaving condition (\ref{eq:slaving_ruck}) to get Eq.~(\ref{eq:Uruck2}).
The adhesion and bending contributions favor small and large width $\lambda$, respectively, and the optimal width can be obtained by minimizing the energy (\ref{eq:Uruck2}), leading to
\begin{align}
\lambda & \sim \Delta^{1/3} \ellbc^{2/3}, \label{eq:lambda_ruck_inext} \\
U^\mathrm{ruck} & \sim \Gamma^{2/3} B^{1/3}\Delta^{1/3}\sim YL\beta\epsilon^{1/3}\tDelta^{1/3}. \label{eq:Uruck_opt}
\end{align}
The energy as a function of $\Delta$ is depicted in Fig.~\ref{fig:Energy} (blue curve at small $\Delta$).

The characteristic slope of the ruck is obtained from Eqs.~(\ref{eq:slaving_ruck}, \ref{eq:lambda_ruck_inext}):
\begin{equation}
\frac{A}{\lambda}\sim \left(\frac{\Delta}{\ellbc} \right)^{1/3}
\sim \left(\frac{\tDelta^2}{\epsilon} \right)^{1/6}.
\end{equation}
The small slope assumption is thus valid as long as
\begin{equation}\label{eq:Delta_st}
\tDelta\ll \tDelta^*\sim \epsilon^{1/2},
\end{equation}
which corresponds to $\Delta\ll\ellbc$.
If $\tDelta>\tDelta^*$, the delaminated state is instead a large-slope fold, which we describe below (Fig.~\ref{fig:scheme_main}(b,c)).

The delamination threshold can be obtained by comparing the energies of the ruck and of the laminated state (Eqs.~(\ref{eq:Ulam}, \ref{eq:Uruck_opt})), yielding
\begin{equation}
\tDelta_\textrm{d} \sim \left(\beta^3\epsilon\right)^{1/5}.
\end{equation}
This estimate is valid only if the small-slope approximation is valid at delamination, $\tDelta_\textrm{d}<\tDelta^*$, which corresponds to moderate bendability:
\begin{equation}\label{eq:validity_ss}
\epsilon\gg\beta^2.
\end{equation}
In this regime, upon increasing the confinement $\tDelta$, the sheet delaminates at $\tDelta_\textrm{d}$ into a small-slope ruck, which then becomes steeper and finally reaches self-contact and becomes a large-slope fold at $\tDelta^*$; this is the phenomenology described quantitatively by Wagner and Vella~\cite{Wagner2013} and Napoli and Turzi~\cite{Napoli2017}.
On the contrary, at high bendability, $\epsilon\ll\beta^2$, the small-slope approximation is not valid at delamination and the critical confinement $\tDelta_\textrm{d}$ should be determined by comparing the energy of the planar state to the energy of a large-slope fold.

Finally, we can evaluate the residual compressive stress in the ruck from
\begin{equation}\label{eq:stress_ruck}
\sigma_\mathrm{ruck}=-\frac{\partial U^\mathrm{ruck}}{\partial\Delta}\sim \Gamma^{2/3}B^{1/3}\Delta^{-2/3}.
\end{equation}
Using the value of $\lambda$ (Eq.~(\ref{eq:lambda_ruck_inext})), the stress can be written
\begin{equation}
\sigma_\mathrm{ruck}\sim \frac{B}{\lambda^2},
\end{equation}
which corresponds to the critical load in Euler buckling and wrinkling instabilities.

\subsection{Large-slope fold}
\label{subsec:fold}

For a confinement above the threshold value $\tDelta^*\sim \epsilon^{1/2}$ (Eq.~(\ref{eq:Delta_st})), that is $\Delta^*\sim \ellbc$, where the small slope assumption is not valid, we can devise a fold ansatz that can absorb any excess length at constant energy, in the spirit of the large folds occuring in floating sheets at large confinement (Fig.~\ref{fig:scheme_main}(b))~\cite{Demery2014d}.
In such a fold the adhesion and bending energies are concentrated in the delaminated region of size $\lambda$ and at the tip of the fold, with size $\lambda'$.
There is no bending energy in the ``walls'' of the fold and, due to self-contact, there is no adhesion energy either.
Since the lengths $\lambda$ and $\lambda'$ are determined by a balance of adhesion and bending, they are governed solely by the bendo-capillary length: $\lambda\sim\lambda'\sim\ellbc$.
Note that the tip of the fold corresponds to the ``racket'' shape described earlier \cite{Cohen2003, Py2007}.
Hence, the delaminated length and the energy of the fold are
\begin{align}
\lambda&\sim\ellbc,\\
U^\mathrm{fold}&\sim \sqrt{B\Gamma}\sim YL\beta\epsilon^{1/2},\label{eq:Ufold}
\end{align}
and the fold amplitude is trivially determined by a slaving
condition, akin to Eq.~(\ref{eq:slaving_ruck}):
\begin{equation}\label{eq:slaving_fold}
A\simeq \frac{\Delta}{2}.
\end{equation}
At the ruck-fold transition, $\Delta\sim\ellbc$, the ruck and fold energies (Eqs.~(\ref{eq:Uruck_opt}, \ref{eq:Ufold})) are equal, signifying yet another transition, separate from the delamination transition.
In the fold state the stress vanishes, as is evident from the independence of the energy on the confinement:
\begin{equation}
\sigma_\mathrm{fold}=-\frac{\partial U^\mathrm{fold}}{\partial\Delta}=0.
\end{equation}
The energy plateau attained upon the formation of a fold is shown in Fig.~\ref{fig:Energy} (blue curve at large $\Delta$).

The delamination threshold in the high bendability regime $\epsilon\ll\beta^2$ is obtained by comparing the energy of the fold to the energy of the laminated state (Eqs.~(\ref{eq:Ufold}, \ref{eq:Ulam}), respectively), leading to 
\begin{equation}
\tDelta_\textrm{d} \sim \beta^{1/2}\epsilon^{1/4}.
\end{equation}
In this regime, the flat state delaminates into a large-slope fold, whose amplitude grows proportionally to the confinement $\tDelta$~(Fig.~\ref{fig:scheme_main}(c)).

\section{Compressible ruck, bifurcation diagram}
\label{sec:Bifurcation}

The scaling analysis of the previous section and the resulting phase diagram (Fig.~\ref{fig:scheme_main}) along with the discontinuous transitions it embodies raise some natural questions: to what extent is the inextensibility assumption, which we employed to describe the delaminated states, a valid one? 
What are the energy barriers associated with the transitions between the different states? 
How is delamination affected by defects, such as patches of the substrate without adhesion?
To address these questions, we consider the energy (\ref{eq:Utot}) when compression is allowed, such that $U_\mathrm{strain}>0$ also in the delaminated state.
First, this allows us to assess the validity of the inextensibility assumption underlying the analysis in the previous section.
Second, delaminated states with compression are capable of interpolating between the laminated and the different delaminated states, thus enabling us to evaluate the energy barriers underlying the delamination transition, from which we can deduce the effect of defects.

We start with a thorough analysis of the small slope limit.
We write the energy of a small-slope ruck with compression allowed.
Then, optimizing the energy with respect to the amplitude $A$ for a given width $\lambda$ of the delaminated zone, we obtain a bifurcation diagram that we compare to the inextensible case.
Finally, we consider the energy in the presence of a defect and show how the behavior of the system can be extracted from the bifurcation diagram.

\subsection{Energy of a ruck in a compressible sheet}\label{sec:rescaling}

We consider the energy (\ref{eq:Uruck}) of a small-slope ruck, allowing for compression and thereby replacing the slaving condition (\ref{eq:slaving_ruck}) by a strain energy:
\begin{align}
U & = U_\mathrm{ad}+U_\mathrm{bend}+U_\mathrm{strain}\\
& = \Gamma\lambda + \frac{c_1 BA^2}{2\lambda^3}+\frac{YL}{2}\left(\tilde\Delta-\frac{c_2A^2}{\lambda L} \right)^2,\label{eq:Uruck_comp}
\end{align}
The numerical constants $c_1$ and $c_2$ can be determined by matching to the small slope \emph{elastica} solution~\cite{Wagner2013},
\begin{equation}
h(x)=\frac{A}{2}\left[1+\cos \left(\frac{2\pi x}{\lambda} \right) \right]
\end{equation}
for $x\in[-\lambda/2,\lambda/2]$, leading to
\begin{align}
c_1 & = 2\pi^4,\\
c_2 & = \frac{\pi^2}{4}.
\end{align}
Finally, it is more convenient to consider directly the energy difference with respect to the laminated state:
\begin{equation}
\Delta U=\frac{c_1 BA^2}{2\lambda^3}+\Gamma\lambda+YL\left[\frac{1}{2}\left(\frac{c_2 A^2}{\lambda L}\right)^2-\frac{c_2 A^2 \tilde\Delta}{\lambda L}\right].
\end{equation}

We point out that the laminated state is a limit case of a compressible ruck corresponding to $A=0$, $\lambda\to 0$.
Hence delamination into a small-slope ruck can be understood by considering the evolution of the energy landscape $\Delta U(A,\lambda)$ as the confinement $\tDelta$ increases, assuming fixed values of $\beta$ and $\epsilon$.

Introducing the dimensionless parameter
\begin{equation}
\eta = \frac{c_1\beta^3\epsilon}{2c_2}
\end{equation}
and the rescaled versions of the energy $u$, confinement $\delta$, ruck's width $x$ and amplitude $a$
\begin{align}
U & = YL\eta^{2/5} u,\\
\Delta & = L\eta^{1/5}\delta,\\
\lambda & = L\eta^{2/5}\beta^{-1} x,\\
A & = c_2^{-1/2} L\beta^{-1/2}\eta^{3/10} a,
\end{align}
the energy difference reads
\begin{equation}\label{eq:energy_diff_rescaled}
\Delta u = x+\left(\frac{1}{x^2}-\delta \right)\frac{a^2}{x}+\frac{a^4}{2x^2}.
\end{equation}

Notably, the confinement $\delta$ is the only parameter left after rescaling, thereby drastically simplifying the study of the energy landscape $\Delta u(a,x)$.
This implies, in particular, that delamination occurs at a critical value $\delta_\textrm{d}\sim 1$, meaning that $\tDelta_\textrm{d}\sim \eta^{1/5}\sim \left(\beta^3\epsilon \right)^{1/5}$.
This observation substantiates the validity of the scaling of the threshold obtained in Sec.~\ref{sub:scaling_ruck} by assuming that the ruck is inextensible; we see below that the presence of compressive strain in the ruck does however affect the numerical prefactors associated with this scaling relation.

The rescaling also provides the characteristic slope of the ruck: assuming that $x\sim a\sim 1$, we get $A/\lambda\sim (\beta^2/\epsilon)^{1/10}$, thereby substantiating the validity of the small slope approximation in the parameter regime $\epsilon\gg\beta^2$ (Eq.~(\ref{eq:validity_ss})), which was obtained in Sec.~\ref{sec:Inextent} by assuming an inextensible ruck.

\subsection{Optimal amplitude}\label{}

The energy landscape $\Delta u(a,x)$ has a simple form: for a given value of $x$ it is a second order polynomial in $a^2$, meaning that it has a single minimum $a^*(x)$.
We thus find the optimal amplitude of a ruck for a given value of its width $x$.
If 
\begin{equation}\label{eq:condition_x_eb}
x\geq x_\textrm{E}=\delta^{-1/2},
\end{equation}
$\Delta u$ is minimal for the amplitude
\begin{equation}\label{eq:opt_amplitude}
a^* = \sqrt{\delta x-\frac{1}{x}}.
\end{equation}
With this amplitude the energy is
\begin{equation}\label{eq:energy_rescaled_opt}
\Delta u = x-\frac{1}{2}\left(\delta-\frac{1}{x^2} \right)^2.
\end{equation}
Otherwise, if $x<x_\textrm{E}$, the optimal amplitude is $a=0$: the energy of any ruck with a rescaled width below $x_\textrm{E}$ is larger than the energy of the fully laminated state.

The threshold value $x_\textrm{E}$ (Eq.~(\ref{eq:condition_x_eb})) of the ruck's width $x$ at a given confinement $\delta$ corresponds to the threshold for Euler buckling.
Indeed, in dimensional quantities, $x>x_\textrm{E}$ reads $\sigma\geq 4\pi^2 B/\lambda^2$, which is the threshold value of the uniaxial compressive load at which a sheet of width $\lambda$ with clamped edges undergoes Euler buckling.
When $x<x_\textrm{E}$, or $\delta < x^{-2}$, the sheet is stable against buckling and the laminated state is favorable even in the absence of adhesion.
Equivalently, for a given confinement $\delta$, the sheet cannot buckle and \emph{a fortiori} delaminate over lengths smaller than $x_\textrm{E}$.
When the Euler buckling threshold is exceeded, the optimal amplitude (\ref{eq:opt_amplitude}) of a ruck of rescaled width $x$ is determined by a balance of strain energy, which favors an amplitude value close to the slaving condition $a=\sqrt{\delta x}$ (Eq.~(\ref{eq:slaving_ruck})) to relieve the strain of the laminated state, and bending energy, which favors small curvatures and thus small amplitudes. 
The optimal amplitude departs continuously from $0$ as a function of $x$.

As noted above, the strain is not completely relieved with the optimal amplitude (\ref{eq:opt_amplitude}).
We can obtain the residual strain, which underlies the third term of the energy (\ref{eq:Uruck_comp}):
\begin{equation}
\varepsilon_\mathrm{res}=\frac{c_2A^2}{\lambda L}-\tilde\Delta
=\eta^{1/5} \left(\frac{a^2}{x}-\delta \right)
=-\frac{\eta^{1/5} }{x^2}=-\frac{4\pi^2 B}{Y\lambda^2}.
\end{equation}
As we noted above, this corresponds to the critical Euler strain, which remains in a finite-amplitude ruck.

\subsection{Inextensibility limit}\label{}

\begin{figure}
\begin{center}
\includegraphics[scale=1]{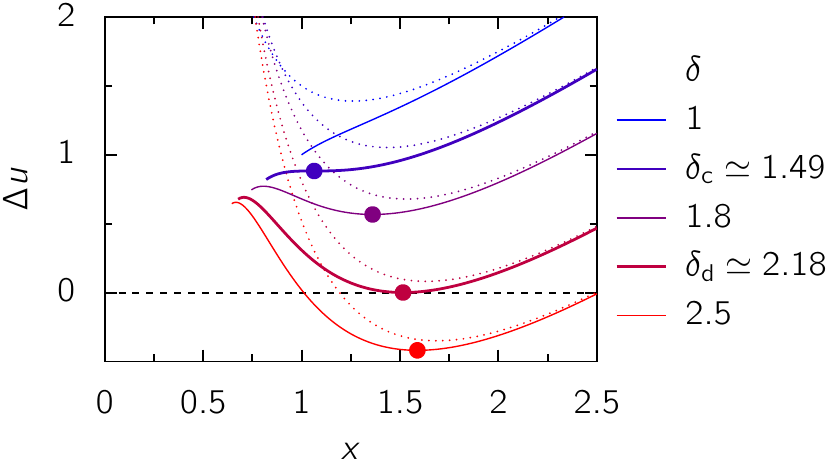}
\end{center}
\caption{Energy difference $\Delta u$ as a function of the rescaled ruck's width $x$ with the amplitude $a$ chosen to minimize the energy (Eq.~(\ref{eq:energy_rescaled_opt})) and local minimum (circles) for different values of $\delta$.
Dotted lines: energy $\Delta u^\mathrm{inext}$ obtained upon modelling the delaminated state as an inextensible deformation of the sheet.}  
\label{fig:lambda_U}
\end{figure}

Before addressing the general solution of Eqs.~(\ref{eq:opt_amplitude}, \ref{eq:energy_rescaled_opt}), let us reconsider the inextensible limit.
Assuming that the delaminated sheet is inextensible amounts to satisfying the slaving condition (\ref{eq:slaving_ruck}), which reads $a=\sqrt{\delta x}$ in rescaled form, whereby the strain is completely relieved.
In this case the energy (\ref{eq:energy_diff_rescaled}) is given by
\begin{equation}\label{eq:energy_diff_inext}
\Delta u^\mathrm{inext}=x+\frac{\delta}{x^2}-\frac{\delta^2}{2}.
\end{equation}
Note that for any $x$ this energy is higher than the energy (\ref{eq:energy_rescaled_opt}) obtained with the optimal amplitude, meaning that the actual delaminated state retains some, albeit minute level of residual strain.
The energy $\Delta u^\mathrm{inext}$ is plotted as a function of $x$ for different values of $\delta$ in Fig.~\ref{fig:lambda_U} (dotted lines).
Mimizing this energy over $x$, we obtain
\begin{align}\label{eq:x_inext}
x^\mathrm{inext}&=(2\delta)^{1/3}\\
\Delta u^\mathrm{inext}&=\frac{3}{2^{2/3}}\delta^{1/3}-\frac{\delta^2}{2},
\end{align}
thereby recovering the scaling relations obtained in Sec.~\ref{sub:scaling_ruck}, supplementing them with numerical prefactors.
Delamination occurs when $\Delta u^\mathrm{inext} =0$:
\begin{equation}\label{eq:delta_d_inext}
\delta_{\textrm{d,inext}} = 54^{1/5}\simeq 2.22.
\end{equation}

\subsection{Bifurcation diagram}\label{}

We now analyze the evolution of the energy (\ref{eq:energy_rescaled_opt}) as the confinement increases; we plot it as a function of $x$ for different values of $\delta$ in Fig.~\ref{fig:lambda_U} (solid lines).
At small $\delta$, the energy has a single minimum at $x=x_\textrm{E}=\delta^{-1/2}$, meaning that $a=0$.
As $\delta$ is increased beyond a first threshold
\begin{equation}
\delta_\textrm{c}\simeq 1.49,
\end{equation}
a local maximum $x_\mathrm{max}$ and a local minimum $x^*$ appear, both of which correspond to delaminated states with positive amplitudes.
However, the energy of the new local minimum is still higher than the energy of the laminated state.
As $\delta$ is increased beyond a second threshold
\begin{equation}
\delta_\textrm{d}\simeq 2.18,
\end{equation}
the energy of the local minimum becomes lower than the energy of the laminated state and delamination occurs.
As expected, the threshold $\delta_\textrm{d}$ is lower than the one obtained by assuming inextensibility (Eq.~(\ref{eq:delta_d_inext})), but the relative difference is only about $2 \%$.

\begin{figure}
\begin{center}
\includegraphics[scale=1]{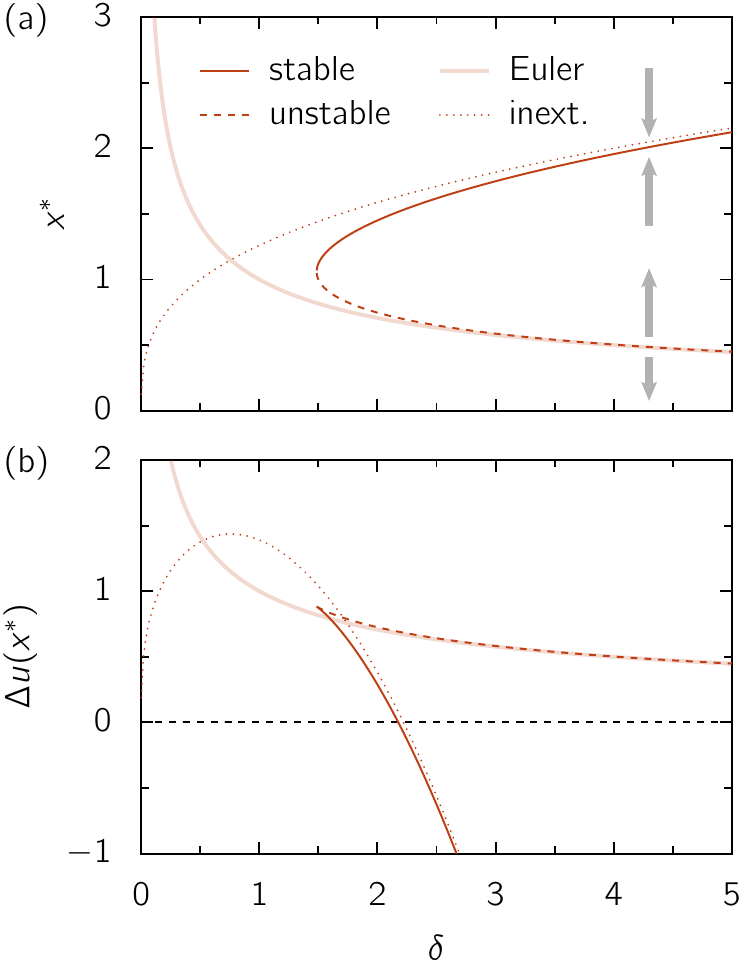}
\end{center}
\caption{(a) Bifurcation diagram that describes delamination in the moderate bendability regime, $\beta^2 \ll \epsilon$. The curves correspond to stable (solid line) and unstable (dashed line) branches, inextensible case (dotted line) and Euler threshold (thick light line).
(b) Energy difference for the local minimum (solid line), the local maximum (dashed line), which corresponds to the energy barrier, in the inextensible case (dotted line), and for the adhesion energy in Euler buckling (thick light line).
}
\label{fig:bifurcation}
\end{figure}

The bifurcation diagram, which consists of stable and unstable branches (local minimum $x^*$ and maximum $x_\mathrm{max}$, respectively) as a function of $\delta$ is shown in Fig.~\ref{fig:bifurcation}(a).
At large confinement, the stable branch approaches the inextensible approximation (Eq.~(\ref{eq:x_inext})), while the unstable branch approaches the threshold for Euler buckling (Eq.~(\ref{eq:condition_x_eb})).
The energy differences between the two branches and the fully laminated state are shown in Fig.~\ref{fig:bifurcation}(b).
Recalling that a discontinuous transition requires the crossing of an energy barrier from the metastable laminated state to the stable delaminated state, we can estimate this gap as $\Delta u(x_\mathrm{max})$ (dashed line in Fig.~\ref{fig:bifurcation}(b).
At large confinement, we find that $x_\mathrm{max}\to x_\textrm{E}=\delta^{-1/2}$: the energy barrier is dominated by the adhesion energy of the corresponding unstable delaminated state.
Notably, the laminated state remains metastable even under arbitrarily large confinement.
Nevertheless, the energy barrier for delamination decreases as $\Delta u\sim \delta^{-1/2}$.

Similarly to other nucleation phenomena (e.g. condensation in a supersaturated gas), which are typically facilitated by the presence of defects rather than by rare thermal fluctuations that enable crossing the relevant energy barrier, we anticipate that the primary mechanism for delamination under compression is also mediated by defects.
This motivates our study in the next section.

\section{Defect mediated delamination}\label{sec:defect}

In this section we consider the nucleation of a delaminated state on a defect, which suppresses the energy barrier.
Inspired by Velankar {\emph{et al.}} \cite{Velankar2012}, we model a defect by a patch of size $\lambdadef$ where there is no adhesion.
Such a defect could originate, for instance, from a gas bubble between the sheet and the substrate, or from a heterogeneous substrate-liquid surface energy.

\subsection{Nucleation of a small-slope ruck}\label{}

We first consider the nucleation of a small-slope ruck, which is relevant for the moderate bendability regime, $\epsilon\gg\beta^2$, using the rescaling introduced in Sec.~\ref{sec:rescaling} and the rescaled defect size $x_\mathrm{def}$.
For a given width $x$ of the delamination zone, the optimal amplitude $a^*$ is determined by a balance of strain and bending energies, regardless of the presence of a defect, so that the defect only affects the adhesion term in the energy (\ref{eq:energy_rescaled_opt}), leading to
\begin{equation}
\Delta u = \max(x-x_\mathrm{def},0)-\left(\delta-\frac{1}{x^2} \right)^2.
\end{equation}

\begin{figure*}
\begin{center}
\includegraphics[width=\linewidth]{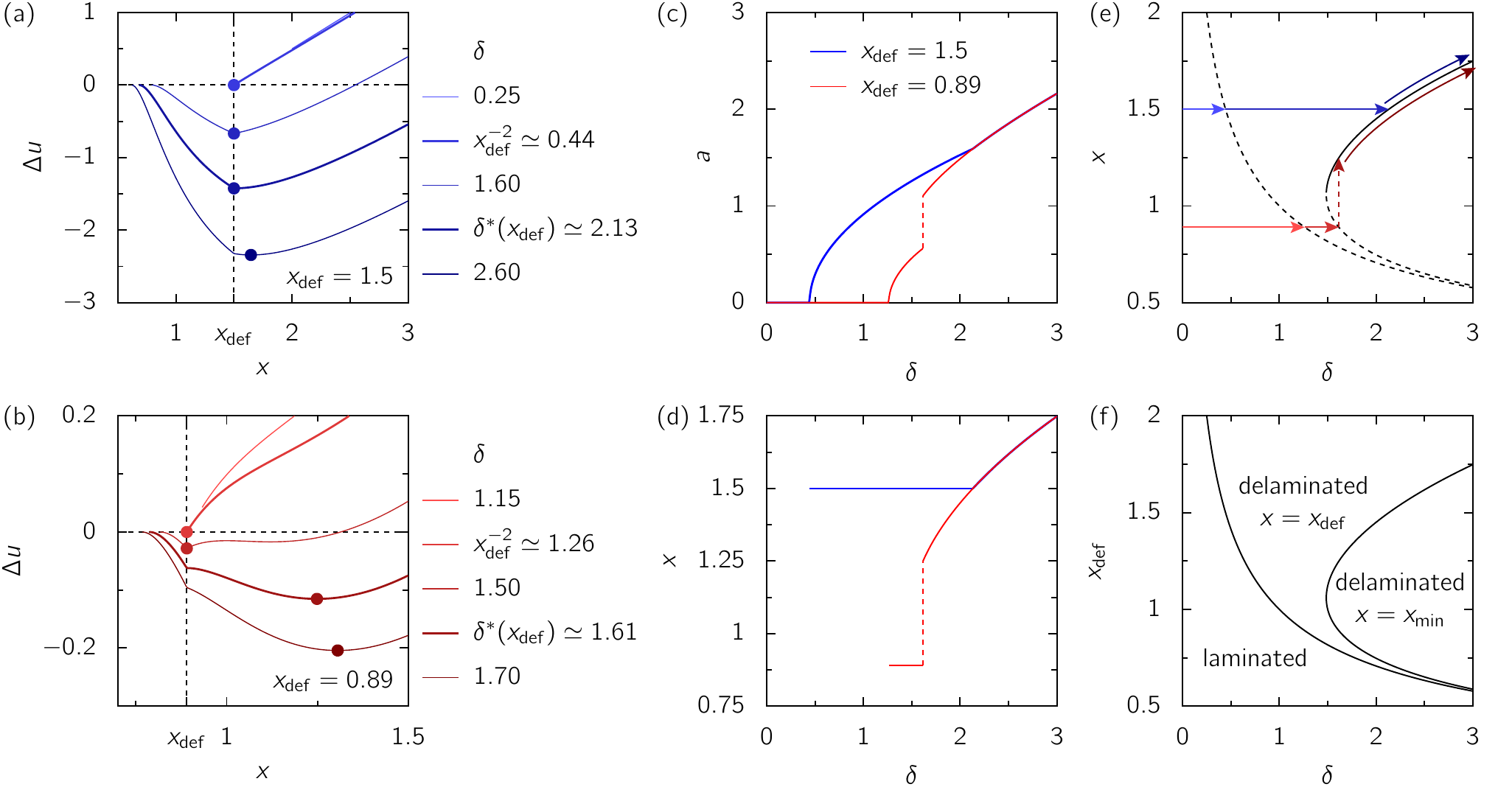}
\end{center}
\caption{The effect of defects on the formation and growth of rucks. The various curves describe the behavior in the presence of a defect of rescaled size $x_\mathrm{def}$; 
$x_\mathrm{def}=0.89$ (red) or $x_\mathrm{def}=1.5$ (blue).
(a,b) Energy difference with the laminated state for different values of $\delta$ (labels) for $x_\mathrm{def}=1.5$ (a) or $x_\mathrm{def}=0.89$ (b).
The circles indicate the state of the system by following the local energy minimum from the laminated state.
(c) Amplitude as a function of $\delta$.
(d) Delaminated width as a function of $\delta$.
(e) Path followed in the two cases in the $(\delta,x)$ plane showing the Euler buckling threshold and bifurcation lines.
(f) Phase diagram summarizing the effect of confinement and defect size on the width of the delaminated zone in various regions of the $(\delta, x_\mathrm{def})$ plane for moderately bendable sheets ($\beta^2 \ll \epsilon$).
}
\label{fig:defect}
\end{figure*}

The energy is plotted as a function of $x$ for different values of $\delta$ and two defect sizes in Fig.~\ref{fig:defect}(a,b).
Starting in the laminated state and following the energy minimum as $\delta$ increases, we see that delamination occurs over the defect when the threshold for Euler buckling is reached: $\delta=x_\mathrm{def}^{-2}$.
The delaminated length $x$ remains set by $x_\mathrm{def}$ and the amplitude grows continuously (Fig.~\ref{fig:defect}(c,d)) until the bifurcation curve is reached (Fig.~\ref{fig:defect}(e)).
The behavior depends on the ratio $x_\mathrm{def}/x_\textrm{c}$, where $x_\textrm{c}\simeq 1.06$ is the width of the delaminated zone at the critical confinement in the defect-free system:
\begin{itemize}
\item If $x_\mathrm{def}>x_\textrm{c}$, the bifurcation curve is met on the stable branch (blue curves in Figs.~\ref{fig:defect}(a,e)), and the energy minimum departs continuously from $x_\textrm{def}$ to follow the stable branch (Fig.~\ref{fig:defect}(d)); the amplitude is continuous (Fig.~\ref{fig:defect}(c)).
\item If $x_\mathrm{def}<x_\textrm{c}$, the bifurcation curve is met on the unstable branch (red curves Fig.~\ref{fig:defect}(b,e)), and the energy minimum jumps from $x_\mathrm{def}$ to the stable branch (Fig.~\ref{fig:defect}(d)); the amplitude is discontinuous (Fig.~\ref{fig:defect}(c)).
In this case the behavior is hysteretic and the system would not follow the same path upon decreasing $\delta$: it would follow the stable branch of the bifurcation curve down to the critical point and the delaminated length would then jump down to the defect size, $x=x_\mathrm{def}$.
\end{itemize}

The effect of the defect size $x_\mathrm{def}$ and confinement $\delta$ on the width $x$ of the delaminated zone can be summarized in a phase diagram (Fig.~\ref{fig:defect}(f)).
Below the Euler buckling threshold, $\delta\leq x_\mathrm{def}^{-2}$, the sheet is laminated.
Between this line and the bifurcation curve, the sheet delaminates and the delamination width is given by the size of the defect, $x=x_\mathrm{def}$. 
In the bifurcation curve, the delamination width is given by the stable branch of the bifurcation curve, $x=x^*$.

\subsection{Delamination into a fold}\label{}

As we saw in Sec.~\ref{sec:Inextent}, at large bendability, $\epsilon\ll\beta^2$, the sheet adopts a folded configuration right at delamination. 
Contrary to the moderate bendability regime, we do not have a finite-strain ansatz that interpolates between the compressed laminated state and the inextensible fold state from which we could derive the bifurcation diagram.
However, using elementary considerations about the large-slope fold and the analysis of the nucleation in a small-slope ruck we can propose a bifurcation diagram also in the large bendability regime.

\begin{figure}
\begin{center}
\includegraphics[scale=1]{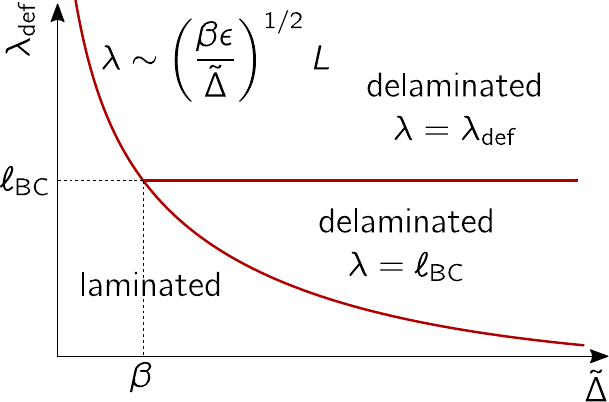}
\end{center}
\caption{Phase diagram analogous to Fig.~\ref{fig:defect}(f), but for the high bendability regime, $\epsilon\ll\beta^2$, where the shape that emerges at delamination threshold is a fold, rather than a small amplitude ruck.}
\label{fig:defect_fold}
\end{figure}

We focus on adapting the diagram of Fig.~\ref{fig:defect}(e) to delamination into a large-slope fold.
Ignoring the residual strain in the fold state, the curves in the $(\delta, x_\mathrm{def})$ plane that are derived from the unstable and stable branches of delamination are given, respectively, by the Euler buckling criterion, $x_\textrm{E}=\delta^{-1/2}$ (such that an infinitesimal-amplitude delamination of width $x>x_\textrm{E}$ gives rise to a fold), and the inextensible fold state addressed in Sec.~\ref{sec:Inextent}, which is characterized by a width $\lambda=\ellbc$. 
These considerations readily yield the phase diagram shown in Fig.~\ref{fig:defect_fold} (in the original version of the parameters).

\section{Beyond wet adhesion on a rigid substrate}
\label{sec:beyond}

\subsection{Dry adhesion}
\label{sec:dry}

\begin{figure}
\begin{center}
\includegraphics[scale=1]{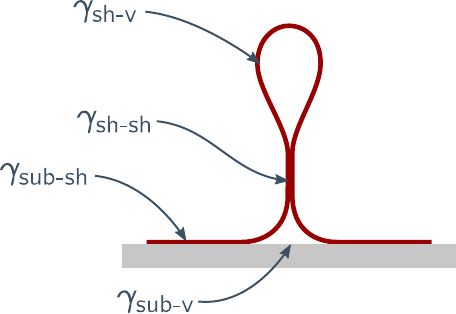}
\end{center}
\caption{Surface energies involved in dry adhesion}
\label{fig:dry_energies}
\end{figure}

We have focused on wet adhesion, where the unique contribution to adhesion energy comes from the formation of liquid-vapor interfaces upon delamination ($\Gamma=2\glv$).
In the absence of a lubricating liquid film (``dry adhesion"), a proper modelling of the compressed laminated state and the delamination process may require one to account also for tangential forces between the sheet and the substrate, whose nature may be conservative \cite{Kothari2020} or frictional \cite{Kolinski2009, Vella2009}. Notwithstanding these complications, it is useful to discuss the necessary modification of the adhesion energy (\ref{eq:Utot}) and its direct implications. In the case of dry adhesion four surfaces energies are involved: substrate-vapor $\gamma_\textrm{sub-v}$, sheet-vapor $\gamma_\textrm{sh-v}$, substrate-sheet $\gamma_\textrm{sub-sh}$, and sheet-sheet $\gamma_\textrm{sh-sh}$ for the self-contact in the fold (Fig.~\ref{fig:dry_energies}).
A natural assumption is that $\gamma_\textrm{sh-sh} = 0$ or merely much smaller than the other 3 surface energies, however we will consider also the possibility that all 4 surface energies are comparable.

When the delaminated shape is a small-slope ruck, the picture of wet adhesion holds, with the adhesion energy replaced by $\Gamma=\gamma_\textrm{sub-v}+\gamma_\textrm{sh-v}-\gamma_\textrm{sub-sh}$.

The large-slope fold is affected differently.
The size of the racket at the tip of the fold is set by the bendo-capillary length $\ellbc'=\sqrt{B/\Gamma'}$, where $\Gamma'=\gamma_\textrm{sh-v}-\gamma_\textrm{sh-sh}$.
The length of the delaminated area at the base of the fold is affected by $\ellbc=\sqrt{B/\Gamma}$ as well as $\ellbc'$.
The most significant difference between the dry adhesion problem and its wet counterpart pertains to the energy of the fold, which now has a contribution proportional to its amplitude coming from self-contacting parts,
\begin{equation}\label{eq:Ufold_dry}
U^\textrm{fold}\sim \Delta\gamma_\textrm{fold}\sim 2A\gamma_\textrm{fold}
\end{equation}
per unit length in the transverse direction, with
\begin{equation}
\gamma_\textrm{fold}=Y\beta_\textrm{fold}=\frac{1}{2}\gamma_\textrm{sh-sh}-\gamma_\textrm{sub-sh}.
\end{equation}
This additional contribution modifies the threshold for fold formation and we may distinguish between two cases.
\begin{itemize}
\item $\gamma_\textrm{fold}<0$: In this case (which is the ultimate one if $\gamma_\textrm{sh-sh}\approx 0$), the sheet prefers to stick to itself rather than to stick to the substrate, so that there is an energy gain in increasing the fold amplitude. 
For highly bendable sheets ($\epsilon\ll\beta^2$), delamination occurs when the energy of the fold becomes negative:
\begin{equation}
\tDelta_\textrm{d}\sim \epsilon^{1/2} \ , 
\end{equation}
such that the delamination threshold is now smaller than the threshold in the wet adhesion case (\ref{eq:VHB}). 
\item $\gamma_\textrm{fold}>0$: In this case (which is possible only if $\gamma_\textrm{sh-sh}$ exceeds twice the value of 
$\gamma_\textrm{sub-sh}$), there is an energetic penalty for growing the fold.  
Comparing the energy (\ref{eq:Ufold_dry}) to the energy of the laminated state, we obtain the new delamination threshold
\begin{equation}
\tDelta_\textrm{d}\sim\beta_\textrm{fold} \ . 
\end{equation}
For $\epsilon\ll\beta^2$, this threshold is higher than the threshold in the wet adhesion case (Eq.~(\ref{eq:VHB})).
\end{itemize}

\subsection{Tangential sheet-substrate forces}\label{}

We expect a similar demarcation of high and low bendability regimes that exhibit, respectively, small-slope rucks and large-slope folds at the delamination transition, to be relevant also for other examples of thin sheets that deflect from a rigid substrate under uniaxial compression. 
One example is a recent study \cite{Kothari2020}, in which the effect of conservative (non-frictional) tangential forces exerted by the substrate on the laminated parts of the sheet was modelled 
through an additional contribution to the energy (\ref{eq:Utot}) that is harmonic in the horizontal (in-plane) displacement and proportional to a substrate-sheet ``bond stretching'' constant $k$. The presence of such a term yields a new length scale, $\sqrt{Y/k} = \sqrt{E t/k}$, over which the stress induced by a compressive load on each edge decays. Hence, one may distinguish two cases:
(a) If $L \ll \sqrt{Y/k}$, then the effect of this energy term is negligible, and the energy can be safely approximated by Eq.~(\ref{eq:Utot}).
(b) If $L \gg \sqrt{Y/k}$ (which is the case addressed in Ref.~\cite{Kothari2020}), a compressive stress in the laminated state exists only over a length $L_{\rm eff} \sim \sqrt{Y/k}$ and the delamination phenomenology is thus expected to be described by our study (up to different numerical prefactors), upon substitution $\epsilon \to (\ellbc/L_{\rm eff})^2 = E^2t^4/(k \Gamma)$.
The predictions obtained through the analysis in Ref.~\cite{Kothari2020}---delamination into small-slope rucks that evolve to folds (or ``jamming" \cite{Kothari2020}) upon increasing load---resemble the observations in Ref.~\cite{Wagner2013} and hence appear to be specialized to the moderate bendability regime.
We expect a distinct phenomenology, with folds appearing immediately at the delamination transition, to be obtained for sufficiently thin sheets ($t \ll (k\Gamma/E^2)^{1/4}$).

\subsection{Gravity limited deflection from a non-adhesive substrate}\label{}

Let us consider now the ``ruck in a rug'' problem, namely, the deflection of a heavy sheet under uniaxial compression from a non-adhesive substrate.
Here, the mechanical equilibrium of a deflected shape is determined by a balance between gravitational potential energy (GPE, rather than adhesion energy) and bending energy.
Following Ref.~\cite{Kolinski2009} we consider a small-slope ($A \ll \lambda$), inextensible ruck and evaluate its energy, similarly to the  analysis in Sec.~\ref{sec:Inextent}, replacing the adhesion energy ($\Gamma \lambda$) with the GPE, $\rho g t A \sim \rho g t \sqrt{\Delta \lambda}$, where $\rho$ is the mass density of the sheet.
This analysis yields the width of the deflected zone $\lambda \sim \ellbg^{6/7} \Delta^{1/7}$, where $\ellbg = (B/\rho g t)^{1/3}  \sim (E t^2/\rho g)^{1/3}$ is a ``bendo-gravity'' length, akin to the bendo-capillary length (\ref{eq:lbc}), and the amplitude is determined by the inextensibility assumption, $A \sim \sqrt{\Delta \lambda}$.
Comparing the energy of the deflected state with that of the uniformly strained planar state, we find the threshold for ruck formation $\tDelta_\textrm{d} \sim \beta_\textrm{g}^{14/27} \epsilon_\textrm{g}^{1/9}$, where we define the bendability parameter $\epsilon_\textrm{g}^{-1} = (L/\ellbg)^2$, analogously to Eq.~(\ref{eq:epsilon}) and $\beta_\textrm{g} = \rho g t/E$.
The characteristic ruck slope at threshold is $A/\lambda \sim \beta_\textrm{g}^{2/9}/\epsilon_\textrm{g}^{1/6}$, and as it approaches $O(1)$ the ruck assumption becomes invalid.
Hence, we expect to have also for this model system a ``moderate bendability" regime, $\epsilon_\textrm{g}\gg\beta_\textrm{g}^{4/3}$, which features a ruck formation at the deflection threshold, and a ``high bendability'' regime, $\epsilon_\textrm{g} \ll \beta_\textrm{g}^{4/3}$, where a large slope deflection appears already at threshold.
In terms of the dimensional parameters of the model, these two regimes correspond to $ L\ll E/\rho g$ and $ L\gg E/ \rho g$, respectively, independently of the sheet thickness $t$! 

\begin{figure}
\begin{center}
\includegraphics[scale=1]{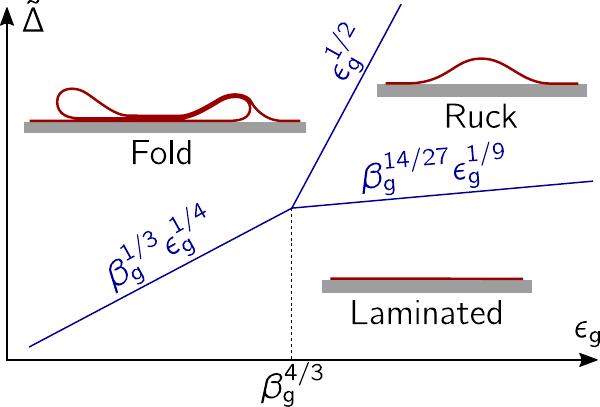}
\caption{Schematic phase diagram for the deflection of a heavy sheet of mass density $\rho$, thickness $t$ and length $L$, under uniaxial compression from a rigid non-adhesive floor, projected onto the plane of dimensionless parameters $\epsilon_\textrm{g} = (\ellbg/L)^2  = (Et^2/\rho g L^3)^{2/3} $ and  $\tDelta)$ for a fixed value of 
$\beta_\textrm{g} = \rho g t/E$. In analogy to Fig.~\ref{fig:scheme_main}(c), the diagram shows the three possible morphologies: flat (uniformly compressed), small-slope ruck and large-slope fold.
}
\label{fig:phase_diag_gravity}
\end{center}
\end{figure}

We note that $E/ \rho g$ is the maximal height, $h_\textrm{max}$, of a solid material (a taller object cannot sustain its own weight). 
Hence, the above observation suggests that upon compressing a solid sheet whose lateral size $L$ exceeds $h_\textrm{max}$, the sheet will fold onto itself at delamination.     
This argument indicates that a plausible fold-like ansatz that minimizes simultaneously bending energy and GPE for $\epsilon_\textrm{g} \ll \beta_\textrm{g}^{4/3}$ can be obtained in analogy to Ref.~\cite{Demery2014d} and the analogous discussion in Sec.~\ref{sec:Inextent}.
As Fig.~\ref{fig:phase_diag_gravity} shows, in such a deflection bending energy and GPE are penalized only at two regions, each of length $\sim \ellbg$ and curvature $\sim \ellbg^{-1}$. Comparing the energy of this fold ansatz with the energy of a flat, uniformly compressed state, we find a deflection transition at $\tDelta_c \sim \beta^{1/3} \epsilon^{1/4}$. The resulting phenomenology is depicted in Fig.~\ref{fig:phase_diag_gravity} (analogously to Fig.~\ref{fig:scheme_main}(b) for the adhesive model). 
In drawing this analogy between delamination from an adhesive (no gravity) and the non-adhesive substrate, we only focus on the energetic mechanism. We cannot rule out the possibility that in the latter case there is an energy barrier or another obstacle to forming a fold.

\subsection{Delamination from a liquid bath}\label{}

While a distinction between low-bendability and high-bendability regimes, analogous to the one highlighted in our paper, may be relevant also for delamination from deformable substrates \cite{Vella2009,Brau2013,Ebata2012,Son2019}, a thorough discussion of the various model systems 
is beyond the scope of our paper. Instead, let us consider only the specific case of a sheet floating on a liquid bath (where the GPE is governed by the liquid density $\rho_\textrm{l}$ rather than the mass of the sheet itself) under uniaxial compression \cite{Pocivavsek2008,Diamant2011,Brau2013}. 
For this system, the planar state may become unstable to wrinkling---periodic undulations of wavelength $\lambda_\textrm{w} \sim (B/\rho_\textrm{l} g)^{1/4}$ (reflecting a balance of bending energy and GPE) that populate the whole sheet without delaminating from the liquid subphase.
In contrast to delamination, wrinkling occurs through a continuous bifurcation, namely, there is no energy barrier and the wrinkle amplitude grows continuously from zero when the confinement $\tDelta$ exceeds the threshold value 
\begin{equation}
    \tDelta_\textrm{w} \sim \frac{\sqrt{B \rho_\textrm{l}g}}{Y}. \label{eq:thre-w}
\end{equation} 
For $\tDelta > \tDelta_\textrm{w}$, the energy of the wrinkled state is
\begin{equation}\label{eq:Uwr}
U^\textrm{wr} \sim \sqrt{B\rho_\textrm{l} g} \tDelta L.
\end{equation} 
On the contrary, if delamination occurs the liquid bath may remain nearly flat with negligible GPE penalty, reflecting a balance of bending and adhesion energies through a single delaminated zone, as described in Sec.~\ref{sec:Inextent}.

Inspection of Eqs.~(\ref{eq:thre-w}, \ref{eq:VHB}, \ref{eq:MHB}) reveals that if the sheet's length $L$ is larger than $L^*$ then the delamination threshold is reached before the wrinkling threshold, $\tDelta_\textrm{d}<\tDelta_\textrm{w}$.
Introducing the ``softness'' parameter \cite{Huang2010}: 
\begin{equation}
s = \frac{\ellbc}{\ell_\textrm{C}} = \frac{\sqrt{B\rho_\textrm{l} g}}{\Gamma} \label{eq:softness} \ ,  
\end{equation}  
where $\ell_\textrm{C} = \sqrt{\Gamma/\rho_\textrm{l} g}$ is the capillary length, we find that:
\begin{equation}
L^*\sim \frac{\ell_\textrm{C}}{\beta}\times\left\{
\begin{array}{lll}
s^{-3/2}&\textrm{if}&s\ll 1,\\
s^{-1}&\textrm{if}&s\gg 1.
\end{array}\right.
\label{eq:Lmax}
\end{equation}
Pocivavsek {\emph{et al.}} \cite{Pocivavsek2008} used a $10~\si{\micro m}$ thick polyester film, for which the softness is $s\simeq 2$ and the critical length is thus $L^*\simeq 1~\si{\kilo m}$.
Huang {\emph{et al.}} \cite{Huang2010} used polystyrene sheets with typical thickness $200~\si{\nano m}$, leading to $s\simeq 10^{-3}$ and $L^*\simeq 500~\si{\kilo m}$.
Obviously, in both cases (in which the sheet's length $L$ is a few centimeters), we have that $L\ll L^*$ and therefore wrinkling is expected before delamination.

The above consideration suggests that in order to determine the delamination threshold, the energy of the delaminated states (Eqs.~(\ref{eq:Uruck}, \ref{eq:Ufold})) should be compared to the energy of the wrinkled state (Eq.~(\ref{eq:Uwr})) instead of the flat state.
We find that the wrinkling-delamination threshold $\tDelta_\textrm{w-d}$ is governed by the softness $s$:
\begin{equation}
  \tDelta_\textrm{w-d} \sim \frac{\ell_\textrm{C}}{L} \times \left\{ 
  \begin{array}{llll}
    1 & \textrm{if} & s \ll 1 & \textrm{(fold),} \\
    s^{-1/2} & \textrm{if} & s \gg 1 & \textrm{(ruck).}
  \end{array}
  \right.
  \label{eq:wrink-del}
\end{equation}
In this equation we did not take into consideration the wrinkle-fold transition of the \emph{fully laminated} state, which is known to suppress the energy considerably \cite{Pocivavsek2008,Diamant2011,Audoly2011,Demery2014d}.
This transition takes place at $\tDelta\sim \ell_\textrm{C}\sqrt{s}/L$, so that if $s\ll 1$ it occurs before delamination into a fold. 
On the contrary, if $s\gg 1$, delamination into a ruck precedes the wrinkle-fold transition.
As a conclusion, depending on softness parameter $s$, the following sequences are predicted:
\begin{equation}
  \textrm{flat} \to \text{wrinkles} \to \left\{ 
  \begin{array}{ll}
    \textrm{laminated fold} & (s\ll 1), \\
    \textrm{delaminated ruck} & (s\gg 1).
  \end{array}
  \right.
  \label{eq:wrinkles_fold_delam_seq}
\end{equation}

Notably, the prediction (\ref{eq:wrinkles_fold_delam_seq}) does not correspond to the experimental observations.
On one hand, Pocivavsek {\emph{et al.}} ($s\gtrsim 1$) observed a wrinkle-fold transition of the fully laminated state and have not reported a delamination~\cite{Pocivavsek2008}.
This apparent ``paradox'' may be readily resolved by recalling the profound difference between the wrinkle-fold and delamination transitions.
While the former constitutes a continuous cross-over of the laminated shape, the delamination transition has been shown to be sub-critical over a rigid substrate, and the same can be expected for delamination from a wrinkled state over a liquid substrate.
The wrinkled state would thus remain stable under small deflections, being destablized only with the aid of deflections that are sufficiently large to cross the corresponding energy barrier (Fig.~\ref{fig:bifurcation}).
Thus, while we expect that delamination may occur if the confinement is done sufficiently slowly \cite{Wagner2011} (or perhaps with the aid of defects, such as those addressed in Sec.~\ref{sec:defect}), it seems plausible that a typical experiment avoids delamination and thus leads to the wrinkle-fold transition of the laminated shape.

On the other hand, in experiments with ultrathin polymer sheets ($s\ll 1$), Huang \emph{et al.}~\cite{Huang2010} have not observed a wrinkle-fold transition of the fully laminated state.
The avoidance of the wrinkle-fold transition has been attributed to a suppressing effect of a wrinkle cascade, which appears precisely when $s\ll 1$~\cite{Huang2010}.
In that case, Eq.~(\ref{eq:wrink-del}) predicts that the wrinkled state could delaminate into a single fold when $\Delta >\ell_\textrm{C}$, which is certainly attainable in that experimental set-up. 
Although observation of such a macrocopic delamination of the wrinkle pattern may be challenging due to the sub-critical nature of the instability and the associated energy barrier, we note that in the experiments of \cite{Huang2010} it has been observed that the wrinkled sheet often develops ``accordions" of large-slope, tightly packed undulations next to the confining walls. It may be interesting to explore whether this phenomenon reflects a ``localized delamination'' that does not expand throughout the whole length of the sheet.

\section{Discussion}
\label{sec:discussion}

In this article we revisited the delamination of a thin sheet under uniaxial compression from a rigid adhesive substrate. Following previous works on this problem~\cite{Kolinski2009, Wagner2013, Napoli2017}, we employed a basic model energy (\ref{eq:Utot}), which accounts for the adhesion energy, as well as the bending and strain in the sheet. 
Our analysis highlights the important effect of the ratio $\ellbc/L$, between the bendo-capillary length and the length of the confined sheet, on the equilibrium configurations (Figs.~\ref{fig:scheme_main},\ref{fig:Energy}) as well as the kinetics of the delamination transition and the effect of defects (Figs.~\ref{fig:defect}(f),\ref{fig:defect_fold}).
As a consequence, an exhaustive description of delamination transitions predicted by this model energy requires one to specify two dimensionless material parameters---the extensibility $\beta$ and bendability $\epsilon^{-1} = (L/\ellbc)^2$. 
In a hindsight, we note that previous works 
have implicitly addressed a  moderate bendability parameter regime, $\beta^2 \ll \epsilon$, in which the sheet delaminates into a small-slope ruck that evolves into a large-slope fold upon increasing confinement. The current work reveals a distinct, high bendability regime, $\epsilon \ll \beta^2$, in which the confined laminated sheet transforms into a fold immediately at the delamination transition. 

In this regard, it is interesting to consider here the experiments of Velankar {\emph{et al.}} \cite{Velankar2012} on the swelling of polymer (PDMS) films attached to a rigid substrate.
These authors observed that upon swelling the homogeneous state does not develop small-slope ``buckles'', but rather highly-localized, large-slope folds, which are often (but not always) preceded by a creasing instability of the surface.
Velankar {\emph{et al.}} further argued that, rather than merely being a surface instability, the folds are in fact delaminated zones. Since the compressive load in such swelling experiments is biaxial, and the observed patterns consist of multiple delaminated zones, rather than a single fold expected under uniaxial compression, any attempt to test our predictions against the observations of Velankar {\emph{et al.}} \cite{Velankar2012} (or \cite{Singamaneni2010, Ortiz2010}) must be taken with a grain of salt. 
Nonetheless, let us note that with the physical parameters of their system: $L \simeq 25~\si{\milli m}$, $t \simeq 100~\si{\micro m}$, $E\simeq 20~\si{\kilo\pascal}$, and assuming $\Gamma \simeq 0.1~\si{\newton/m}$, we have that $L/t \simeq 250$, whereas $Et/\Gamma \simeq 2$, so that a fold-like delamination instability of the homogeneous state is consistent with our prediction. 
We hope that controlled experiments under uniaxial compressive load will elucidate the effect of the bendability parameter on the nature of delamination patterns.

Thus, while we focused our study on delamination induced by uniaxial compression, we expect that the conceptual lesson drawn from our study, namely, the relevance of a bendability parameter to delamination phenomenology, applies also for other, more complicated systems, specifically the biaxial stresses associated with swelling or thermal expansion of polymer sheets on rigid \cite{Velankar2012,Ortiz2010,Singamaneni2010} and deformable \cite{Ebata2012,Son2019} substrates.
Another notable example is the curvature-induced delamination of sheets from an adhesive substrate of a spherical shape, where compression is induced by geometrical incompatibility \cite{Majidi08,Hure11,Hure2013}.
For that problem too we find that previous studies have focused on a moderate bendability regime and a different approach is required to describe delamination patterns in a high bendability regime.

\begin{acknowledgements}
We thank D. Vella for many discussions and invaluable comments and D. Vella and O. Oshri for their critical reading of this manuscript.
BD acknowledges support by the National Science Foundation under grant DMR 1822439. 
\end{acknowledgements}

\end{document}